\documentstyle[epsf]{mn}
\def\mg2{Mg$_2$}

\begin{document}

\title{Galactic Bulges from HST-NICMOS Observations: Ages and Dust}

\author[Peletier et al.]{Reynier F.~Peletier,$^{1,2,3}$
Marc~Balcells,$^{4}$
Roger L.~Davies,$^{1}$
Y.~Andredakis,$^{2}$
  \newauthor 
A.~Vazdekis,$^{1,5}$
A.~Burkert,$^{6}$
and F.~Prada$^{7,8}$\\
$^{1}$ Dept. of Physics, 
University of Durham, South Road,
Durham, DH1 3LE, UK \\
$^{2}$ Kapteyn Astronomical Institute,
Postbus 800,
9700 AV~~Groningen, The Netherlands \\
$^{3}$ School of Physics and Astronomy, University of Nottingham, 
Nottingham NG7 2RD, UK \\
$^{4}$ Instituto de Astrof\'\i sica de Canarias,
E-38200 La Laguna,
Tenerife, Spain \\
$^{5}$ Institute of Astronomy,
University of Tokyo,
Tokyo, Japan \\
$^{6}$ 
MPI f\"ur Astronomie,
K\"onigstuhl 17,
69117 Heidelberg, Germany \\
$^{7}$ Instituto de Astronom\'\i a, UNAM, Apdo. Postal 877, 22800
Ensenada, B.C. Mexico \\
$^{8}$ Calar Alto Observatory, Almeria, Spain } 

\maketitle

\markboth{R.F.~Peletier et al.: Galactic Bulges from HST-NICMOS Observations:
Ages and Dust}{~}

\begin{abstract}
We present optical and near-infrared colour maps of the central
regions of bulges of S0 and spiral galaxies obtained with WFPC2 and
NICMOS on the HST.  By combined use of HST and ground-based data the colour 
information spans from a few tens of pc to a few kpc.  In almost
all galaxies the colour profiles in the central 100-200 pc become
more rapidly redder.  We attribute the high central colour indices to
a central concentration of dust.  We infer an average extinction at the
center of A$_V$ = 0.6 -- 1.0 mag.  Several objects show central dust
rings or disks at sub-kpc scales similar to those found by others in
giant ellipticals.  For galactic bulges of types S0 to Sb, the
tightness of the $B-I$ vs $I-H$ relation suggests that the age spread
among bulges of early type galaxies is small, of at most 2 Gyr.  Colours
at 1 R$_{\rm eff}$, where we expect extinction to be negligible are 
similar to than those of elliptical galaxies in the Coma 
cluster, suggesting that these bulges formed at the same time as the 
bright galaxies in Coma. Furthermore the galaxy ages are found to be 
independent of their environment. Since it is likely that Coma 
was formed at redshift z $>$ 3, 
our bulges, which are in groups and in the field, also must have been 
formed at this epoch. Bulges of early-type spirals cannot be
formed by secular evolution of
bars at recent epochs, since such bulges would be much
younger. There are three galaxies of type Sbc and later, 
their bulges are younger and could perhaps arise from secular evolution
of transient bars.  
Our results are in good agreement with semi-analytic predictions 
(Baugh {\it et al.} 1996, 1998, Kauffmann 1996), who also predict that bulges,
in clusters and in the field, are as old as giant ellipticals in clusters.

\end{abstract}

\begin{keywords}
~~
\end{keywords}

\section{Introduction}

We do not have a clear picture of the formation mechanism for the
central bulges of spiral galaxies.  Their structural and dynamical
properties have long suggested the view that bulges are like small
elliptical components residing in the center of a large disk.  Indeed,
bulges obey the D$_n$ -- $\sigma$ relation and fall on the Fundamental
Plane of elliptical galaxies (Dressler {\it et al.} 1987,
Bender et al. 1992).  Their surface
brightness profiles fall off as $\mu(r) ~\propto~ r^{1/n}$ (Andredakis
{\it et al.} 1995), with $n$ = 4 for early-type galaxies
falling to $n$ = 1 for Sc and later types.
Dynamically, bulges are consistent with oblate, isotropic
models, like low-luminosity ellipticals (Kormendy \& Illingworth 1982,
Davies {\it et al.~} (1983)).  In a series of papers Peletier and
Balcells have studied optical and near-infrared (NIR) colour
distributions of a complete sample of spiral bulges, using ground-based
data (Balcells \& Peletier 1994, Peletier \& Balcells 1996, 1997) with
the goal of extending the comparison of bulges and ellipticals to their
stellar populations. They find that, once dust is accounted for, the
colours of bulges are never redder than those of elliptical galaxies of
the same luminosity.  Bulge colours are very similar to those of the
inner disk, the differences being much smaller than colour differences
from galaxy to galaxy.  Population models then suggest that the inner
disk (at 2 scale lengths) must have formed at the same time as, or at
most 3 Gyr after, the bulge, and that bulge metallicities are lower
than those of giant ellipticals.

HST allows us to investigate the centres of bulges with a ten fold
increase in angular resolution, yet HST data on the colours of galaxy
centres at scales of tens of parsecs is scarce.  Colours contain useful
information on the stellar populations of the galaxy centres, and allow
us to estimate the level of internal extinction.  HST data have shown
that dust patches are very common in galaxy nuclei, even in ellipticals
(van Dokkum \& Franx 1995), hence we expect a strong signature of dust
in spiral bulges.  NICMOS allows an unprecedented look at the inner
structure of bulges in the H band where the extinction is approximately
a factor 6 lower than in V. NICMOS can produce more reliable
measurements of surface brightness profiles and isophotal shapes.

Here we show the results of the first colour study of the centers of 
galactic bulges based on HST/NICMOS data.
We observed in the NICMOS F160W 
band ($H$-band), F814W ($I$) and in F450W, a wider version of the $B$-band. 
This combination of passbands is very suitable to study stellar populations
in galaxies with limited amounts of extinction and recent star formation,
like centres of early-type spirals. In those galaxies the observed colours 
are determined by the underlying old stellar population, perturbed by
dust extinction and the light of young, recently formed stars. Extinction
by dust affects all optical and near-infrared colours, while recent star
formation can only be seen in blue passbands (see e.g. Knapen et al. 1995).
In that paper it is shown that a red colour, like $I-H$ will primarily
show the extinction due to dust, as well as old stellar population gradients,
while a colour which includes a blue band like $B$ or $U$, in this paper 
$B-I$, is specially sensitive to recent star formation.
In the absence of extinction, B, I and H make it
possible, in principle, to measure the age and metallicity of the
stellar population (e.g. Aaronson et al. 1978; Peletier et al.  1990; 
Bothun \& Gregg 1990). We infer large amounts of
centrally concentrated dust within the inner 100 pc.  Outside this
region, the colour-colour distribution is quite tight, allowing us to
place limits on the age dispersion of galactic bulges.
With the HST it is easier to reach a high photometric 
accuracy, because there is no atmospheric extinction to correct for,
the instrumental PSF is stable, and the instrument is very well 
characterized, as compared to ground-based instruments. This means that
we can expect accuracies of 0.02--0.03 mag or better 
in optical and near-infrared
colours. Such high quality measurements are vital if we are to separate
the effects of age and metallicity using colours, even though the 
large wavelength baseline ($B$-$I$-$H$) minimises the effect of photometric 
errors.

Currently there are three main scenarios for the formation of bulges: 
bulges form before, contemporaneously with, or after
disks (see the review by Wyse {\it et al.} 1997, and Bouwens {\it et al.} 
1999).  
In the first scenario (Eggen et al. 1962, Larson 1975, Carlberg 1984) 
the formation of bulges is mainly
described by collapse of a primordial gas cloud into clumps, which then merge
together. The disk only forms after the last massive  merger via gas infall. In
a second scenario, infall of a gas-rich dwarf galaxy onto a disk
produces star formation more or less at the same time in bulge and disk 
(Pfenniger 1992). In the
third scenario of secular evolution of disks (e.g Combes {\it et al.} 1990,
Pfenniger \& Norman 1990, Norman {\it et al.} 1996), the bulge is formed by dynamical
instabilities of the disk, in which a bar is formed, which then forms a massive
central concentration by allowing material to stream towards the central
regions. The central  mass concentration itself then tidally disrupts the bar
and forms the bulge. When material is continuing to fall onto the disk sometime
later a new bar  can be formed, and the whole process will start again and
induce more star  formation in the bulge. To distinguish between these models
one needs to measure the ages of both bulges and disks. Peletier \& Balcells
(1996) and Terndrup {\it et al.} 1994, 
using ground-based data, found that the ages of bulges and inner
disks are very similar. Here in this paper we measure the ages of  the bulges
themselves. 
 
The paper is organized as follows: Section 2 gives details of the
observations and data reduction. Section 3a discusses the
dust content of the galactic bulges. Section 3b describes the analysis of the
stellar populations, and in Section 4 some implications for 
our understanding of galactic bulges are discussed.

\section{Sample and Observations}

Twenty galaxies were observed with HST in Cycle 7 (Summer 1997) with WFPC2
(F450W and F814W) and NICMOS (F160W). They all are part of the 
original sample of Balcells \& Peletier (1994), which is a complete,
$B$-magnitude-limited sample of early-type spirals (type S0-Sbc)
with inclinations larger than 50$^{\rm o}$, and for which one side of the 
minor axis colour profile is approximately featureless as seen from the ground. 
The subsample considered here was chosen to include galaxies of  
types S0-Sbc, excluding galaxies that are exactly edge-on, and is biased 
towards the nearest objects. The galaxies observed are given in Tables 1 and 2.
Their basic parameters: luminosities, effective radii and ground-based 
colours, are given in Andredakis {\it et al.~} (1995) and Peletier \& Balcells (1997).
The mean absolute $R$-band magnitude of the sample is
$M_R$ = --22.50 mag, with an RMS dispersion of 1 mag 
(H$_0$ = 70 km~s$^{-1}$~Mpc$^{-1}$).

The standard HST-pipeline data reduction was used for the WFPC2 optical data. 
For NGC~7457 the $I$-band observations were taken from the HST Archive.
For the $H$-band two MULTIACCUM NIC2-exposures of in total 256 s were taken, 
offset from each other by 1$''$, to enable us to remove bad pixels. 
The standard STSDAS CALNICA reduction package was
used, after which a small fraction of the flatfield was subtracted from the 
reduced images, to account for a non-zero pedestal level in the raw image.
This fraction was determined by requiring that the final image was smooth, and
amounts to about 0.3-0.5 ADU~s$^{-1}$, corresponding to about $H$=17 
mag~(arcsec)$^{-2}$.
Applying this step is only important for the lowest flux levels, but the images
improve considerably in quality (they look like the ground-based images).
The final sky background level was determined by comparing the image with
the ground-based image in the same band. To compare the $H$-band 
data with the ground-based data in $K$, we used the following transformation,
derived from the new GISSEL96 models of Bruzual \& Charlot (see Leitherer 
{\it et al.} 1996):
 $$H-K ~=~ 0.111 (I-K) ~-~ 0.0339$$

The images were calibrated to the STMAG system using Holtzmann {\it et al.~} (1995),
using a constant shift of 0.10 mag to correct to infinite aperture. 
The method described in
the same paper was used to iteratively account for the colour term in 
the F450W filter. The colours were corrected for galactic extinction 
using the new dust maps of Schlegel {\it et al.~} (1998), and the Galactic extinction
law (Rieke \& Lebofsky 1985). After this the bands were K-corrected. This
small correction
(from Persson {\it et al.~} 1979) was $\Delta B$ = -5$z$, $\Delta I$ = --$z$ and
$\Delta H$ = 0. 

To account for the difference in Point Spread Functions (PSF) when determining 
the $X$ -- $Y$ colour maps and profiles, the $X$-band image was convolved with
the  $Y$-band PSF, and the $Y$-band image with the $X$-band PSF. These PSFs
were determined with the TinyTim package (Krist 1992). We then determined minor
axis profiles by averaging azimuthally  in wedges of 22.5$^{\rm o}$, centered
on the $H$-band nucleus. We then combined the surface brightness profiles with
the ground-based profiles to cover the whole spatial range of the bulge, by
shifting the  ground-based profile on top of the HST-profile between about
3$''$ and 6$''$. This way the final profiles have high signal-to-noise
everywhere, and the high-quality photometric accuracy of HST data, for which
absolute and relative accuracies of a few percent can be expected (Colina {\it
et al.}  1998). We then subtracted these wedge profiles in pairs of bands to
obtain colour profiles. In Table 1 we list the galaxies and their colours at  the
center and at one R$_{\rm eff}$.

\begin{table}
\begin{center}
\caption{Basic Parameters}
\begin{tabular}{ccccccc}
\hline
\hline
{\small NGC} & {\small T} & {\small R$_{\rm eff}$} & {\small $B-I_{\rm cen}$} 
& {\small $B-I_{\rm eff}$} & 
{\small $I-H_{\rm cen}$} & {\small $I-H_{\rm eff}$} \\
\hline
  5326  &  1  &    2.16 &  2.510 & 2.089   &  2.117 &	1.796  \\
  5389  &  0  &    4.67 &  2.992 & 2.082   &  2.208 &	 1.823 \\
  5422  & -2  &    4.51 &  2.581 & 2.092   &  2.170 &	 1.809 \\
  5443  &  3  &    4.70 &  2.894 & 1.941   &  2.408 &	 1.554 \\
  5475  &  0  &    2.58 &  2.342 & 2.065   &  2.014 &	 1.774 \\
  5577  &  4  &    4.14 &  1.910 & 1.862   &  1.792 &	 1.665 \\
  5587  &  0  &    2.13 &  2.600 & 1.993   &  2.086 &	 1.840 \\
  5689  &  0  &    3.72 &  2.355 & 2.058   &  2.231 &	 1.816 \\
  5707  &  2  &    3.40 &  2.720 & 2.069   &  2.144 &	 1.722 \\
  5719  &  2  &    3.88 &  3.494 & 2.252   &  2.758 &	 1.972 \\
  5746  &  3  &    14.6 &  4.117 & 2.040   &  2.910 &	 1.749 \\
  5838  & -3  &    7.50 &  2.600 & 2.091   &  2.184 &	 1.767 \\
  5854  & -1  &    6.03 &  1.872 & 1.840   &  1.821 &	 1.659 \\
  5879  &  4  &    2.29 &  2.653 & 1.932   &  2.353 &	 1.704 \\
  5965  &  3  &    8.23 &  2.946 & 2.051   &  2.406 &	 1.755 \\
  6010  &  0  &    2.43 &  2.543 & 2.086   &  2.105 &	 1.843 \\
  6504  &  2  &    4.98 &  2.452 & 2.064   &  2.062 &	 1.764 \\
  7331  &  3  &    5.0 &  2.356 & 2.047   &  1.978 &	 1.766 \\
  7457  & -3  &    8.55 &  1.951 & 1.920   &  1.823 &	 1.620 \\
  7537  &  4  &    1.48 &  2.928 & 1.948   &  2.560 &	 1.969 \\
\\
  4472  &  --   &  --     &   --   & 2.357 & -- & 1.947 \\
\hline
\end{tabular}
\end{center}
{\bf Notes to Table 1:} The morphological types in column (2) are
from the RC3 (de Vaucouleurs et al. 1991). The effective radii of the 
bulges in column (3) in arcseconds are from Andredakis et al. (1995),
except for NGC~7331, where the value is an estimate based on Prada 
et al. (1996). Distances in kpc are given in Andredakis et al. (1995).
Also given are the colours of NGC~4472 at 5$''$ (see text).
\end{table}
 
\section{The $B-I$ vs. $I-H$ colour-colour diagram}

\subsection{Dust in Nuclei of Galactic Bulges}

In Fig.~1 we show intensity-maps of the inner regions of the
galaxies, together with $B-I$ and $I-H$ colour maps with the same scale
and orientation, superimposed on $H$-band contour maps. The maps
indicate the position of dust lanes. 
It is clear from the figure that many, maybe
all, galaxies have red nuclei. To show this more clearly we show all
colour profiles obtained in wedges along the dust free side of the minor
axis in Fig.~2. This is not the first time that red nuclei have been
found in spiral galaxies. For example, in our ground-based data
(Peletier \& Balcells 1997) we show several galaxies with red nuclei.
For that sample however the colour profiles on the least dusty side are
generally featureless at radii larger than 1$''$, showing the same
logarithmic colour gradients throughout the bulge. Inside 1$''$ no
information is available about the colours from ground-based data. The
fact that the PSF of the HST data is well known and stable enables us
to correct the colour-profiles and maps for most of the instrumental PSF
effects. In this way one can measure colours down to radii as small as
$\approx$ 0.10$''$ (the diffraction limit is 0.15$''$). At this
resolution the red nuclei are easy to see, and extended in most of our
sample.

Nuclei can be red because of local dust, large foreground dust lanes,
or because of red stellar populations. One usually can distinguish between
dust and stellar population reddening by looking for patchy structures, since 
they are generally due to dust extinction, or to star formation, in combination
with extinction. In Table~2 we indicate the 
nature of the features that we find in the nuclei.

{\small
\begin{center}
\begin{table}
\caption{Nuclear features}
\begin{tabular}{lcccccl}
\hline
\hline
NGC	&  F. &  Nuc.  &   Blue   & 60mu  & 100mu  &  comment \\
~       &  disk    &   d/p &   Patch &  [Jy]     & [Jy]   &    \\
(1)     &  (2)   & (3)   & (4)  & (5)  & (6) & \\
\hline
5326	  &    -- 	&	       +	&	--    &	  --  &   --  & LGC 361\\
5389	  &    +	&	       +	&	--    &	0.33 & 1.51 & \\
5422	  &    + 	&	       +	&	--    &	  --  &	 --  & LGC 373\\
5443      &    -- 	&	       +	&	--    &	0.33 & 1.48 & LGC 373\\
5475	  &    -- 	&	       +	&	  + &	  --  &	 --  & \\
5577      &    --       &	       +	&	MDSF &	0.58 & 1.94 & LGC 379 \\
5587	  &    -- 	&	       +	&	--    &	0.23 & 0.79 & \\
5689	  &    + 	&	       +	&	--    &	0.46 & 1.27 & LGC 384\\
5707	  &    + 	&	       +	&	--    &	  --  &	 --  & LGC 384\\
5719      &    + 	&	       ?	&	--    &	8.05 & 17.1 & LGC 386\\
5746      &    + 	&	       ?	&	--    &	1.33 & 8.88 & LGC 386\\
5838	  &    -- 	&	       +	&	  + &	0.74 & 1.47 & LGC 392\\
5854	  &    -- 	&	       +	&	  + &	  --  &	 --  & LGC 393\\
5879      &    + 	&	       ?	&	MDSF    &	0.29 & 3.04 & LGC 396 \\
5965      &    + 	&	       +	&	--    &	0.39 & 1.84 & \\
6010	  &    -- 	&	       --	&	--    &	  --  &	 --  & \\
6504	  &    -- 	&	       --	&	--    &	  --  &	 --  & \\
7331      &    + 	&	       ?	&	--    &	23.1 & 81.6 & LGC 459\\
7457	  &    -- 	&	       --	&	--    &	  --  &	 --  & \\
7537      &    --	&	       +	&	MDSF   &	  --  &	 --  &  \\
\hline
\end{tabular}

{\bf Notes to Table 2:} A '+' sign in column (2) indicates the presence of
 a large foreground disk, in column (3) the presence of a red nuclear dust patch
 or lane, and in column (4) the presence of a blue patch near the nucleus. 
IRAS fluxes are presented in column (5) and (6). 'MDSF' in column (4) means 
'mixed dust and star formation' in the whole nuclear area. The LGC number is 
the group designation (by Garcia 1993).
\end{table}
\end{center}
}

We find (Table 2) only three galaxies: NGC 6010, 6504 and 7457, without nuclear
dust features, although in some cases a large foreground dust lane
makes nuclear dust difficult to detect. For these three galaxies we
analysed the isophote shapes more carefully.  Dust patches make galaxy
isophotes irregular in particular the third order Fourier terms C3 and
S3 (Carter 1978) will be non-zero, and generally increasing in
amplitude to the blue. S3 and C3 were found to be significantly
non-zero in both $B$ and $I$ in NGC 6010 and 6504 but not in NGC~7457.
We conclude that the signature of dust is to be found in almost all
galaxies in the sample, including some galaxies indicated with a (--)
sign in column 3 of Table 2.

How much reddening is caused by the dust? In Fig.~3 we show the colours
at the center (filled dots) and at one bulge effective radius (open
dots) in a $I-H$ vs. $B-I$ colour-colour diagram. Fig.~3 (b) and (c)
show that in all cases the galaxy is redder in the centre than at
r$_e$, by sometimes very large amounts. Since the vector indicating
reddening by dust is almost parallel to the the vector indicating
changes in metallicity, it is not possible to say exactly how much of
the reddening is due to extinction. Lower and upper limits to the
internal extinction may be estimated as follows.

A plausible {\it upper limit} to the reddening is derived by assuming
that the galaxy at 1 R$_{\rm eff}$ is dustfree. This is perhaps
reasonable, since at these radii no structure is seen in the colour
maps, and there is a small dispersion in the colour-colour plot.  We
infer that the central $I-H$ and the $B-I$ colours are reddened by an
average of 0.42~$\pm$~0.06 mag, and 0.61~$\pm$~0.11 mag respectively.
A {\it lower limit} to the reddening can be found if one assumes that
the stellar populations are never redder than those of the central
Virgo galaxy NGC~4472.  Using the data at 5$''$ (or $\sim$ 400 pc) from 
Peletier {\it et al.} (1990) for NGC~4472, converted to $B-I$ and $I-H$ 
as described in the Appendix, we find
that the $I-H$ colours would be reddened by on the average
0.37~$\pm$~0.06 mag and $B-I$ by 0.41~$\pm$~0.10 mag.  Using the
Galactic extinction law (Rieke \& Lebofsky 1985) these numbers
correspond to an average internal extinction A$_V$ between 0.89 and
1.01 mag, if $I-H$ is used, and between 0.56 and 0.78 mag when one uses
$B-I$. The estimates using $I-H$ are somewhat higher than those from
$B-I$, because nuclear star formation in some galaxies makes the radial
colour difference in $B-I$ smaller, as is seen in NGC~5838, 5854 and
5475.  In the $B-I$ colour map of these galaxies there are blue nuclear
features which do not appear in $I-H$. These patches, presumably caused
by young stars, reduce the radial colour variations in blue colours
like $B-I$ but not in $I-H$.

Our {\it upper limits} for the amount of dust inferred here are
probably somewhat too high, because part of the reddening is caused by
stellar population gradients. From the most dustfree galaxies we infer
that the colour difference due to stellar populations between center and
1 R$_{\rm eff}$ is $\approx$ 0.1 mag in both $B-I$ and $I-H$,
which implies that colour gradients in the inner parts of bulges are affected
much more by dust than by stellar populations.

\subsection{Stellar Populations}

Apart from providing high resolution, HST also has the advantage that
the photometric conditions are very stable, so that accurate colours can
be determined. For this reason, and because we have a very large
colour-baseline, we can use the colour-colour diagram to infer information
about the age-spread in the sample, and about the cause of the stellar
population gradients.  Studying Fig.~1 carefully, we see that many
physical phenomena are playing a role in these galaxies. We observe the
combined effects of extinction, recent star formation, and old stellar
population gradients. To disentangle them we have displayed the
galaxies in Fig.~1 according to their morphological type. Although the
type (from de Vaucouleurs {\it et al.~} 1991) has been given based on
low-resolution optical observations, we can see that the properties of
the galaxies change smoothly as a function of type. S0 galaxies have
small dustlanes, and have rather featureless colour maps. Sb galaxies
tend to have strong dustlanes, while the Sbc galaxies are considerably
bluer, have lower surface brightness, show patchy dust and star
formation together, and are rather different from the rest of the
galaxies.

Having established (in Section 3.1) that extinction at 1 R$_{\rm eff}$
is probably negligible, we will now consider what the colours can tell us
about the stellar populations. First we note
the small scatter amongst the open symbols in Fig.~3a confirming
that extinction is not important at 1 R$_{\rm eff}$.  If we exclude the
three galaxies with the latest Hubble type (the open crossed symbols in
Fig.~3c) the stellar populations at R$_{\rm eff}$ form a rather tight
sequence in the $B-I$ vs. $I-H$ plane. Fig.~1e shows that the central
regions of our 3 Sbc are full of dust and regions of recent star
formation, we will not consider them further in this analysis of
stellar populations. In Fig.~3a we also show a set of Single Stellar
Population (SSP) models of Vazdekis {\it et al.~} (1996). The models
displayed here have a Salpeter IMF, with a reduced number of low-mass
stars, to match better the Scalo (1986) IMF.  We see that, independent
of the amount of extinction, the tightness of the colour-colour
relation shows that the luminosity weighted age of the stars is very
similar from bulge to bulge. According to the models the age-spread
would be about 1-2 Gyr, although the bluest (and generally faintest)
bulges would be somewhat younger. We derive a mean luminosity weighted
age of 9 Gyr from these models. While the absolute ages are poorly
constrained, the relative ages are much more robust so we can conclude
that the age spread amongst bulges in this sample is small. (If we use
Worthey (1994)'s models the inferred luminosity weighted age would be 
implausibly small : 2 Gyr). The galaxy sequence runs parallel to line
of constant age suggesting that the colour-variations from galaxy to galaxy
are due predominantly to changes in metallicity supporting the view
that the  colour-magnitude relation for early-type galaxies (e.g. Bower {\it et
al.~} 1992) is mainly driven by changes in metallicity.

Figure 3c shows that the type dependence along the colour-colour
relation is very small. The outlying point is NGC~5854 which has a blue
central patch (probably of recent star formation). NGC~7457 is a faint
S0 galaxy that has bluer $B-R$ and $U-R$ colours than an elliptical of
comparable luminosity (Balcells \& Peletier 1994) accounting for its
rather young inferred age. The bulk of the galaxies of type Sb and
earlier occupy a narrow band in $B-I$ vs. $I-H$. As a comparison we
have plotted the colour-colour relation for Coma, determined from the
data of Bower etal (1992), converted to our colours as described in
the Appendix. We find that the colours of our bulges at 1 R$_{\rm eff}$ 
are very similar to those of bright Coma galaxies. Our bulges
however are somewhat bluer in $I-H$ and redder in $B-I$, which, according
to the stellar population models, could be explained if they are 
slightly older and somewhat less metal rich. The colour-colour conversion
is rather uncertain, however. Also plotted in Fig.~3c are 
the colour-colour diagrams converted using the theoretical models of 
Vazdekis {\it et al.} (1996) and Worthey (1994). 
As can be seen in the Figure, the difference between
the three colour-colour diagrams in $B-I$ and $I-K$ is quite large.
Although we argue (see Appendix) that our empirical calibration 
can be trusted much more than the theoretical calibrations, one should
be careful not to over-interpret the data. What seems safe to
conclude is that Coma galaxies are to be found in the shaded area, and
that the ages of our bulges are similar to those of the early-type
galaxies in Coma.

\section{Discussion}

In the previous section we have shown that:
\begin{enumerate}

\item Centers of bulges of early-type spirals are generally dusty.
We find that A$_V$ on the average lies between 0.6 and 1.0 mag, which implies
that A$_H$ should be between 0.1 and 0.2 mag.

\item At 1 R$_{\rm eff}$ Galactic bulges show a very tight $I-H$
vs. $B-I$ relation, implying that the age spread among bulges of
early-type spirals is small (at most 2 Gyr).

\item The colours of bulges of early-type spirals at 1 R$_{\rm eff}$
are similar to the colours of early-type galaxies in the Coma cluster. 
This implies that the age difference between nearby bulges 
and cluster-ellipticals is small, probably smaller than about 2 Gyr.
\end{enumerate}

Since the extinction observed, is more than a factor 2 (0.76 mag), if
the dust is located in or near the center we cannot see through to the
other side of the galaxy. The extinction locally will be very large
(see Sadler \& Gerhard 1985). The fact that this occurs in so many of
our galaxies shows that the very central region is almost always
optically thick in $B$ or $V$.  The situation with dust in the centers
of spiral bulges is similar to that of elliptical galaxies.  Van Dokkum
\& Franx, analyzing WFPC data of nearby elliptical galaxies, found
evidence for central extinction in 75\% of their sample. This is
more than was previously found from the ground (Ebneter {\it et al.~}
1988, V\'eron-Cetty \& V\'eron 1988).  Goudfrooij {\it et al.~} (1994b)
detected dust in 41\% of the galaxies in their sample of Revised
Shapley-Ames (Sandage \& Tammann 1981) galaxies. A similar detection
rate as found from IRAS fluxes by Knapp {\it et al.~} (1989). The
higher angular resolution of HST allows us to reach
much lower detection limits, we have been able to
detect dust in 95 $\pm$ 5\% of our sample of bulges of early-type
spirals. This detection rate is higher than for ellipticals, although
our method of finding dust, using colour images, is more sensitive than
the method of van Dokkum \& Franx, who determined their dust masses
from one $V$-band image only.  Van Dokkum \& Franx found an average
dust mass of 4 $\times$ 10$^3$ M$_\odot$.  For the bulges analysed here
we find an average dust mass of $\sim$ 10$^4$ M$_\odot$, determined 
using the method of van Dokkum \& Franx: for each dust feature the
mass is calculated using M$_{dust}$ = $\Sigma$ $\langle A_V\rangle$ 
$\Gamma_V^{-1}$, with $\Sigma$ the area of the feature, $\langle A_V\rangle$
the mean absorption in the area, and $\Gamma_V$ the visual mass
absorption coefficient (Sadler \& Gerhard 1985). $\Gamma_V$ is taken 
to be 6~10$^{-6}$ mag~kpc$^2$~M$_\odot^{-1}$. Rough values of $\langle A_V\rangle$ 
were determined by taking the difference in $B-I$ or $I-H$ of
the feature and the dustfree values at 1 R$_{\rm eff}$, converted to
A$_V$ using the Galactic extinction law (Rieke \& Lebofsky 1985).  
If we assume a
Galactic gas to dust ratio of 130 (see van Dokkum \& Franx 1995) we
find the galaxies analysed in this paper have an average of 10$^6$
M$_\odot$ of interstellar material in their nuclear regions.

The origin of the nuclear dust is unclear. At larger radii kinematic
observations of gas, usually associated with dust, in ellipticals show
that it is often decoupled from the stellar velocity field (for a
discussion see Goudfrooij {\it et al.~} 1994b).  This is used to imply
that {\it large-scale} dust lanes are of external origin perhaps being
accreted during galaxy mergers or interactions.  However, the origin of
the small arcsecond-scale dust lanes found in ellipticals, often
oriented along the major axis of the stellar body (Goudfrooij {\it et
al.~} 1994b), could well be internal.  Scaling from the numbers for
stellar mass-loss for bright ellipticals given by Faber \& Gallagher,
1976, we estimate that for typical bulges in our sample the mass will
be dposited into the bulge at the rate of $\approx$ 0.1 -- 1
M$_\odot$~yr$^{-1}$, so there is no problem accruing the dust we see.
The amount of dust, the large detected fraction, and the fact that the dust lanes
are found parallel to the major axis indicate an internal origin. The
central dust provides a suitable environment for centrally concentrated
star formation, probably leading to strong metallicity enhancements in
the central 100 pc of bulges. In that regime, it is possible that so
much dust is produced that ordinary dust destruction mechanisms are
ineffective.  Potentially, high resolution spectral observations will
show us whether this dust is indeed of internal origin, and what
metallicities that are being reached. The dust however could provide
some serious observational difficulties for the determination of the
inner slope of the stellar surface density profile of bulges, even in
the H band.

Several of our bulges have central features which resemble the inner
disk in the giant elliptical NGC~4261 (Jaffe {\it et al.~}
1996). Examples are NGC 5326, NGC~5587, NGC~5838 and NGC~5854. NGC~4261
has a LINER spectrum in the center, Jaffe {\it et al.~} argue that this
inner disk might provide the fuel for the AGN.  Only 4 of the objects
studied in our paper however are known to be (weakly) active, NGC 5746,
NGC 5838, NGC 5879 and NGC 7331, all classified by Ho {\it et al.~}
(1997) as T2, transition objects between LINER and Seyfert, with narrow
emission lines. Their colours are however entirely consistent with an
extincted stellar population. Furthermore, many of the ellipticals with
similar features are not active galaxies. It seems that an inner disk
of gas and dust alone is not sufficient to produce an AGN.

No HST colour profiles of bulges of spirals have appeared in the
literature up to now. The only paper presenting high-resolution HST
colour profiles of ellipticals is Carollo {\it et
al.~} (1997a), who report V and I profiles of 15 galaxies with
dynamically decoupled cores using the refurbished WFPC2.  They find
that their galaxies all have very similar $V-I$ gradients between radii
of 1.5$''$ and 10$''$, while the dispersion is larger between 0.25$''$
and 1.5$''$, where some galaxies are seen with gradients that are twice
as large, while some others are much smaller, or even negative. Clearly
the behavior of $V-I$ is different in their inner 1.5$''$ as compared
to the area further out. Although Carollo {\it et al.} (1997a)
masked out dusty areas before obtaining the colour profiles by radially
averaging the remaining light, this process will probably not have removed
all the extinction. The large spread in colour gradient in the inner region
might be a confirmation of significant 
quantities of dust near the center in almost all elliptical galaxies. 

What can we learn about the formation of bulges?
The fact that the ages of most of the bulges in this paper are
so similar and old makes it very difficult to form the bulges of early-type
spirals (S0-Sb) through secular evolution of disks. In this
scenario it is expected that bulges regularly undergo major bursts 
of star formation, to convert gas that has been funneled to the central
area through the presence of a bar into stars. This would mean that
we would expect to find more young bulges and a large spread in bulge ages.
Another problem for early-type bulges is that the bulge densities are up to a
factor 100 higher than in the center of the disks. To
create such high overdensities in the disk, the required star formation
rates are such that they could easily disrupt the disk  (Ostriker 1990).
Late type galaxies (Sbc's and later types) might be different. 
We see that the
three in our sample have younger ages (crossed circles in Fig.~3), and  also
the density contrast between bulge and disk is much smaller. For these galaxies
in the bulge region spiral arms, blue star formation regions  and dust lanes
are seen, as in the disk, telling us that the stellar populations of bulge and
disk here are very much the same (see Fig.~1e). Although  Norman {\it et al.}
(1996) use the fact that the Fundamental Plane has changed little from a
redshift of the order of 0.5 to now as an argument against secular-evolution
driven bulge building, it is not clear whether bulges of late-type galaxies
really lie on  the fundamental plane of ellipticals and early-type bulges,
since no data is available at present.

The fact that the ages of our bulges are all so similar supports the
idea of a major episode of star formation in the past, in which most
of these bulges (and also the bright galaxies in large clusters like
Coma) were formed. It is currently thought that cluster ellipticals
must have formed at redshifts beyond $z$ = 3 (Ellis {\it et al.~}
1997, Stanford {\it et al.~} 1997), because of the lack of evolution
seen in clusters of intermediate redshift. This corresponds to an age
of 10.5 Gyr ($\Omega_{\rm 0}$ = 0.2, $\Omega_\Lambda$ = 0.8, 
H$_{\rm 0}$ = 70 km~s$^{-1}$~Mpc$^{-1}$),
or 8.4 Gyr ($\Omega_{\rm 0}$ = 1, $\Omega_\Lambda$ = 0, 
H$_{\rm 0}$ = 70 km~s$^{-1}$~Mpc$^{-1}$) (Hogg 1999).
The fact that the colours
of the majority of our bulges are similar to the Coma galaxies 
indicates that our bulges (except maybe
the Sbc's) are also old, and formed at redshifts beyond $z$ = 3. These
observations endorse the original model by Eggen {\it et al.}
(1962) forming bulges in the beginning during a monolithic, or clumpy,
collapse. The fact that the colours of bulges and inner disks are very
similar (Terndrup {\it et al.} 1994, Peletier \& Balcells 1996) then
implies that the disk formed very gradually from inside out, with the
age of the inner disk similar to the age of the bulge. Can we then say
something about the age of the disk? Would it be possible that the
whole disk formed at the same time as the bulge? At that point a
conflict arises with the star formation history of the universe
derived e.g. by Madau {\it et al.~} (1996) on the basis of data of the
Hubble Deep Field, which shows a maximum between $z$ = 1 and 3. Since
our bulges are found in a variety of environments, they are in no way
special, and if most early-type spirals would form at redshifts beyond
3 this would also imply that the maximum in the HDF would have to go
to larger z. However, this problem would be solved if disks of
early-type galaxies on the average would be considerably younger than
their bulges. With a sufficiently large age difference between
bulge and disk the luminosity-weighted age of the galaxy (which is 
in general dominated by the disk) can then be young enough.
There is nothing in the data of Terndrup {\it et al.}
1994 and Peletier \& Balcells 1996 preventing this from happening.

Alternatively, Kauffmann (1995) and Governato {\it et al.~} (1998)
have pointed out that in biased models of galaxy formation accelerated
evolution can be expected in dense regions.  Our observations here
show that there is no difference in age between the ellipticals in
Coma and our galactic bulges. This would mean that our bulges, except
for the latest types, must have formed early, beyond $z$ = 3, as well.
On the other hand, we also do not find any dependence of bulge age
upon environment. The bulges are found in all kinds of environments,
ranging from isolated to groups of more than 20 members (see Table
1). Apparently everywhere in the universe these intermediate-size
galaxies must have started forming early-on.

It appears that the only feasible solution is that bulges of early-type
spirals are old, and disks considerably younger. This is in 
agreement with Abraham {\it et. al.~} (1998), who find for the disk 
galaxies in the HDF that their bulges are significantly older than
their disks (up to 50\%). One has however to take into account that
at z $\sim$ 1 many large spiral galaxies are found (Lilly {\it et al.~} (1998), 
whose stellar
populations are consistent with a declining activity since at least 
z=1.5 -- 2 (Abraham {\it et al.~} 1998). 
Semi-analytic galaxy formation models also support this picture.
In Fig.~4 a histogram is shown of the distribution of $V$-band
averaged ages of ellipticals and bulges in the simulations of
Baugh {\it et al.} (1996) (Baugh, private communication). The
models find that the large majority of bulges is old. This is 
independent of the environment, contrary to elliptical galaxies,
which are found to be old in rich clusters, and can be 5 Gyrs 
younger in the field. Bulges are old because the
accompanying disk needs time to form without being disrupted.

We have found that bulges of early-type spirals in general are old, that
their age-spread is smaller than 2 Gyr, and that their colour gradients are 
mainly due to metallicity gradients in the stellar populations. 
These results once more support the idea that
bulges and ellipticals are similar objects. The good agreement
between the observational results of this paper and models like
the semi-analytical models of Baugh {\it et al.} (1996) strongly supports
the picture that bulges were formed through monolithic or clumpy collapse,
and formed much before theier disks. However, one can see that bulges
of Sbc galaxies are different in many respects, and it is likely that these
differences will be larger for Sc galaxies and later types (see also the 
recent HST-study of Carollo {\it et al.} (1997b, 1998). They are smaller,
younger, have lower central surface brightness, and it is not clear
whether they fall on the fundamental plane of elliptical galaxies. Since 
the observations until now are not good enough to establish whether these
late type bulges can be made from bars, and in this way the galaxy type
can be changed by secular evolution, it is clear that it is very important
to study the bulges of late type spirals to understand the formation of bulges.

\section*{Acknowledgements}  

This paper is based on observation with the Hubble Space Telescope
The authors acknowledge very useful support from Luis Colina, 
Massimo Stiavelli, Jeremy Walsh and Doris Daou at the STScI and ST-ECF, 
and useful discussions
with Ian Smail, Harald Kuntschner, Carlton Baugh and Carlos Frenk.
RLD is grateful to Durham University for the award of a Sir James Knott
Fellowship and to the Leverhulme Trust for the award of a Research  
Fellowship, these awards contributed significantly to this research.

\onecolumn
\begin{figure}
\caption{$B-I$ and $I-H$ colour maps, and intensity-maps of the galaxies. 
Greyscale levels for $B-I$ are 1.6,1.7,1.8,1.9,2.0,2.2,2.4,2.7,3.0 and 3.5 mag,
and for $I-H$ 1.6,1.7,1.8,1.9,2.0,2.1,2.2,2.3,2.4 and 3.0 mag. 
Superimposed on the colour maps are $H$-band
contours with levels from 20 to 10 mag with intervals of 1 mag.
The $B-I-H$ maps are black and white representations of 'real colour' maps
made from $B$, $I$ and $H$, and are uncalibrated. 
{\bf a:} S0 galaxies.}
\end{figure}
\addtocounter{figure}{-1}
\begin{figure}
\caption{{\bf b:} S0/a galaxies.}
\end{figure}
\addtocounter{figure}{-1}
\begin{figure}
\caption{{\bf c:} S0/a-Sab galaxies.}
\end{figure}
\addtocounter{figure}{-1}
\begin{figure}
\caption{{\bf d:} Sab-Sb galaxies.}
\end{figure}
\addtocounter{figure}{-1}
\begin{figure}
\caption{{\bf e:} Sb-Sbc galaxies.}
\end{figure}

\begin{figure}
\mbox{\epsfxsize=17cm  \epsfbox{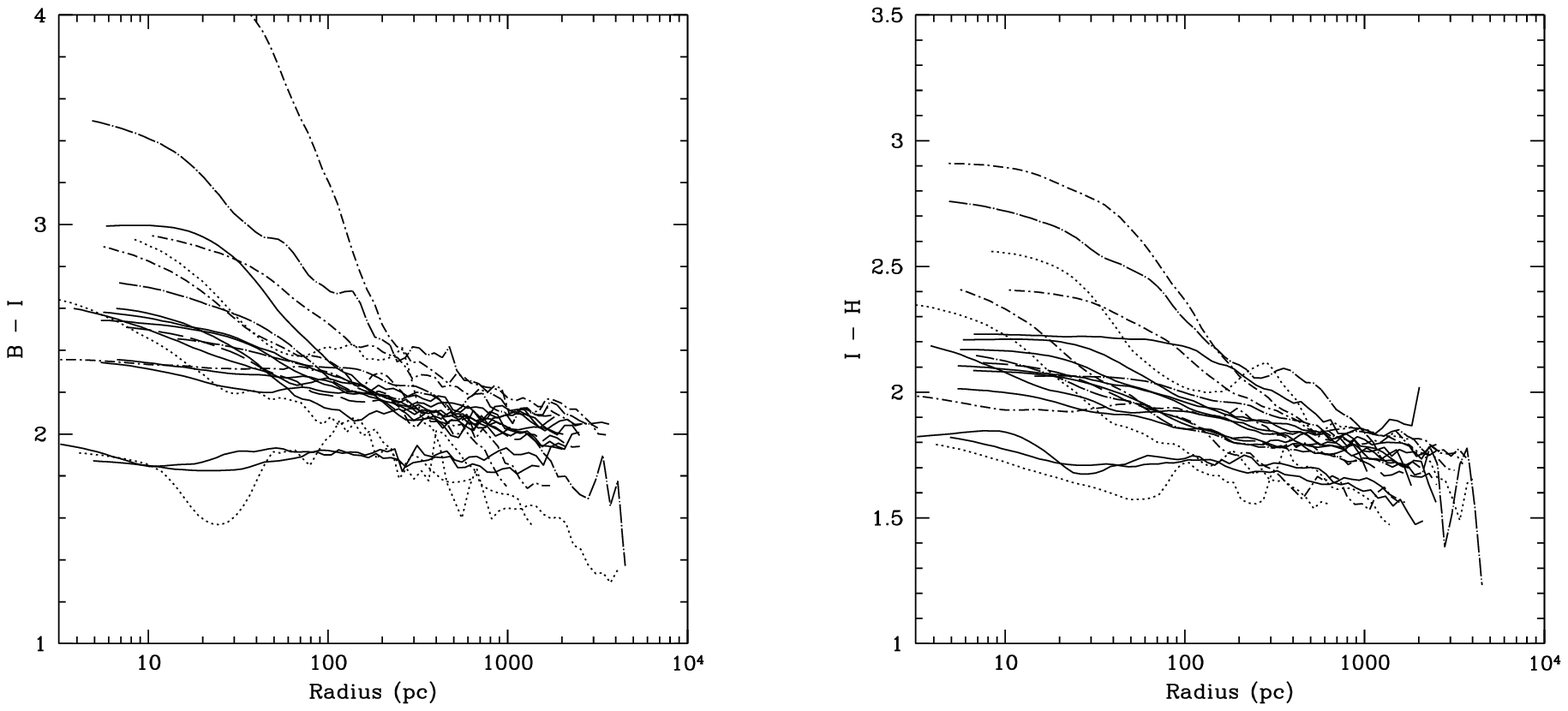}}
\caption{Individual colour profiles for the sample galaxies.
Each galaxy is represented by a line from the center to the point where the 
bulge contribution is as large as that of the disk. Morphological types
are indicated by line types: T $\le$ 0 - drawn lines; T = 1: long gashed;
T = 2: dot - long dashed; T = 3: dot - short dashed; T = 4: dotted. }
\end{figure}

\begin{figure}
\mbox{\epsfxsize=15cm  \epsfbox{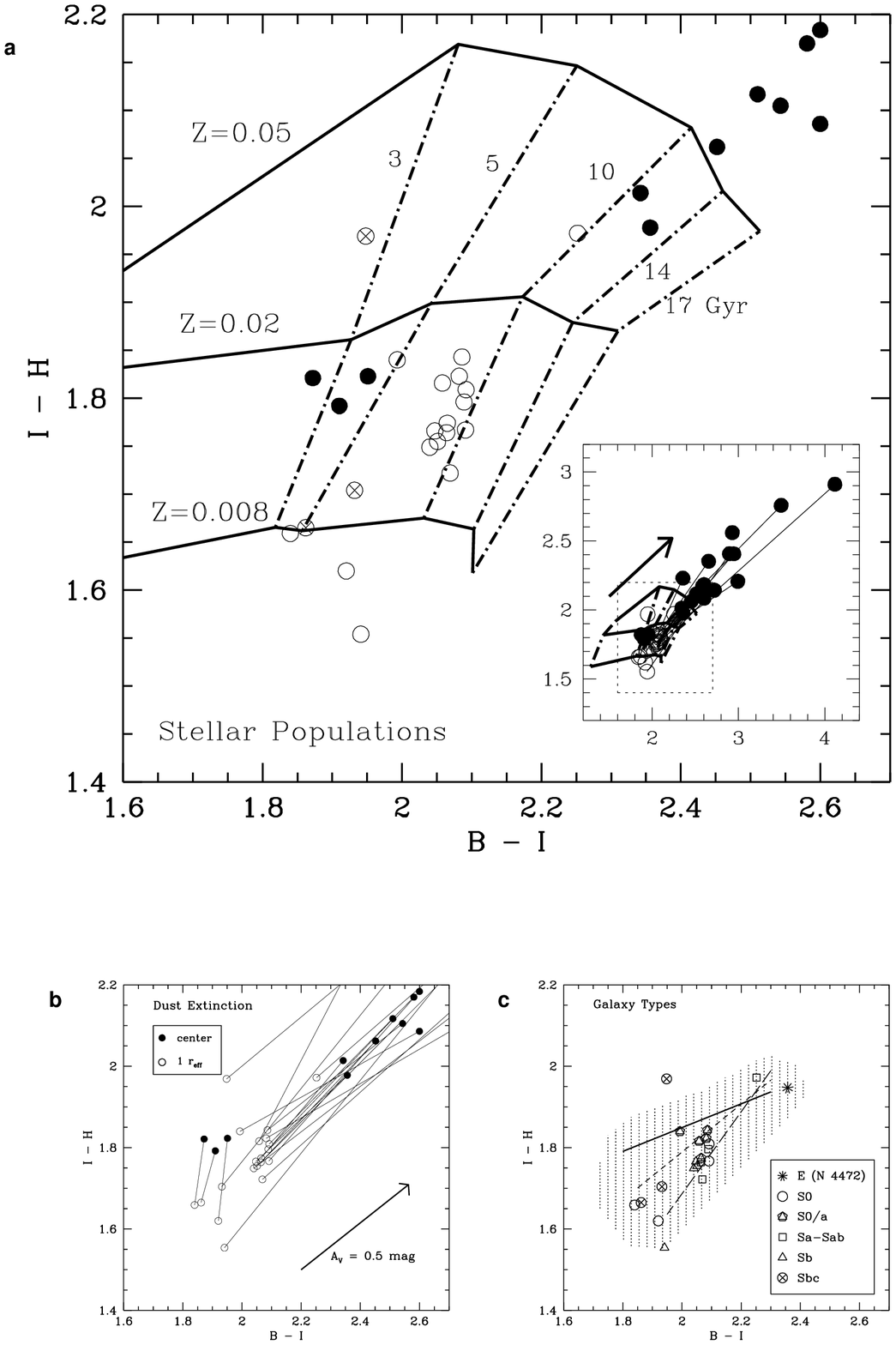}}
\caption{ Colour-colour diagrams for the 20 galaxies. 
In Fig.~3(a) and (b) are displayed the positions of the center (filled) and 
1 bulge effective radius on the minor axis (open circles). 
Reddenings vector for a reddening of $A_V$ = 0.5 mag (3b) and 1 mag (inset
of 3a) are given as well. Superimposed in Fig.~3a are SSP models
by Vazdekis {\it et al.~} (1996). Solid lines are lines of constant metallicity,
dashed-dotted lines are loci of constant age. In Fig.~3c the same 
galaxies at 1 R$_{\rm eff}$ are plotted, with their symbols coded
as a function of morphological type. Also added are NGC~4472.
An attempt has 
been made to convert the data of Bower et al. (1992) to $B-I$ and $I-H$,
using the models of Worthey (1994, long-dashed), Vazdekis et al (short-dashed)
and the empirical calibration described in the Appendix (drawn line)). 
Since these calibrations do not
agree very well with each other we can only say that the early-type
galaxies of Coma are to be found in the shaded area, and that 
the position of
our bulges is consistent with their colours.  }
\end{figure}


\begin{figure}
\mbox{\epsfxsize=15cm  \epsfbox{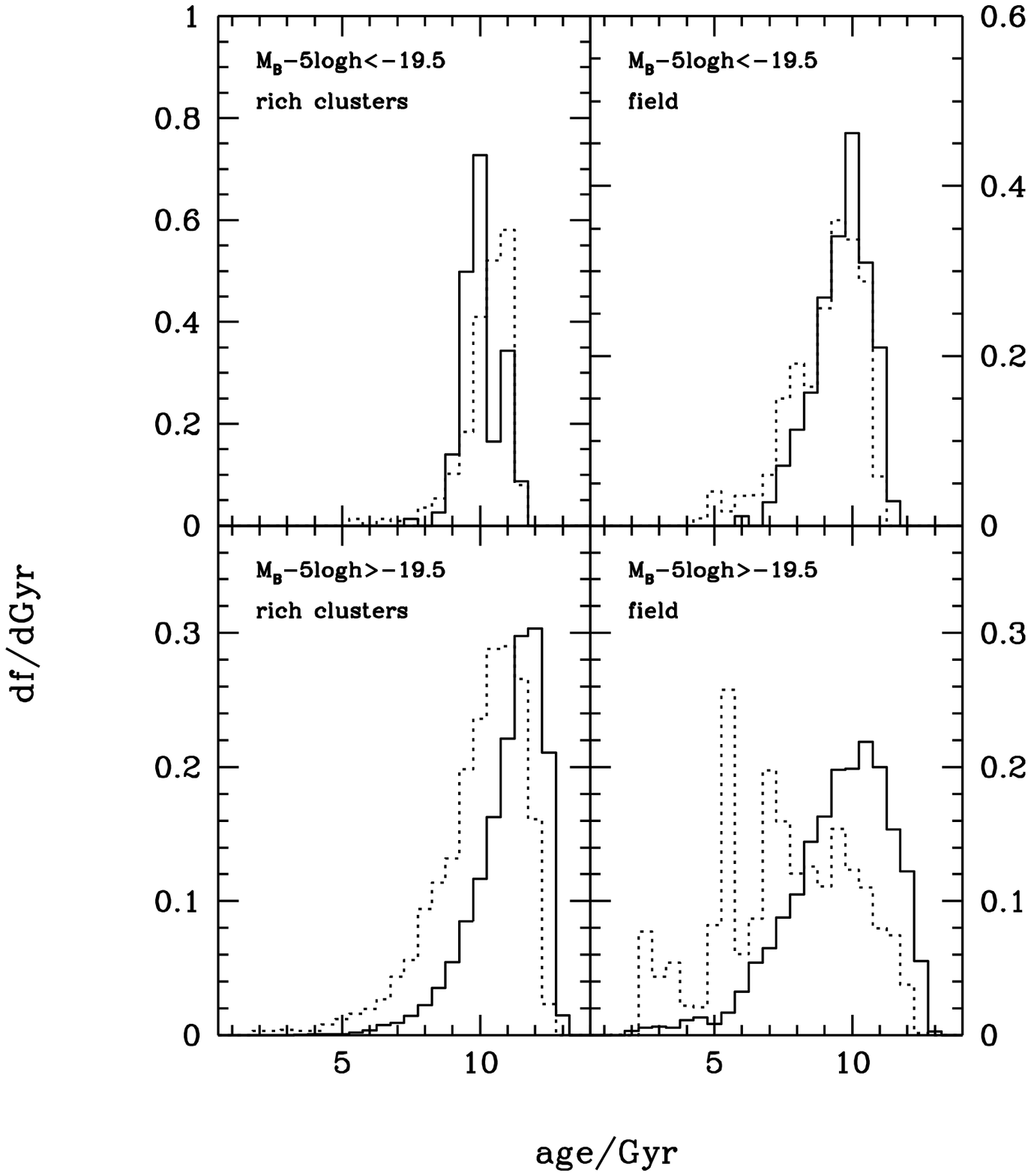}}
\caption{Mean V-band luminosity weighted ages of ellipticals 
and bulges in clusters and in the field, predicted by semi-analytic simulations
by Baugh (private communication), following prescriptions given in Baugh {\it et
al.} (1996).
Drawn lines are bulges, while elliptical galaxies are indicated by dashed lines.
}
\end{figure}
\begin{figure}
\mbox{\epsfxsize=15cm  \epsfbox{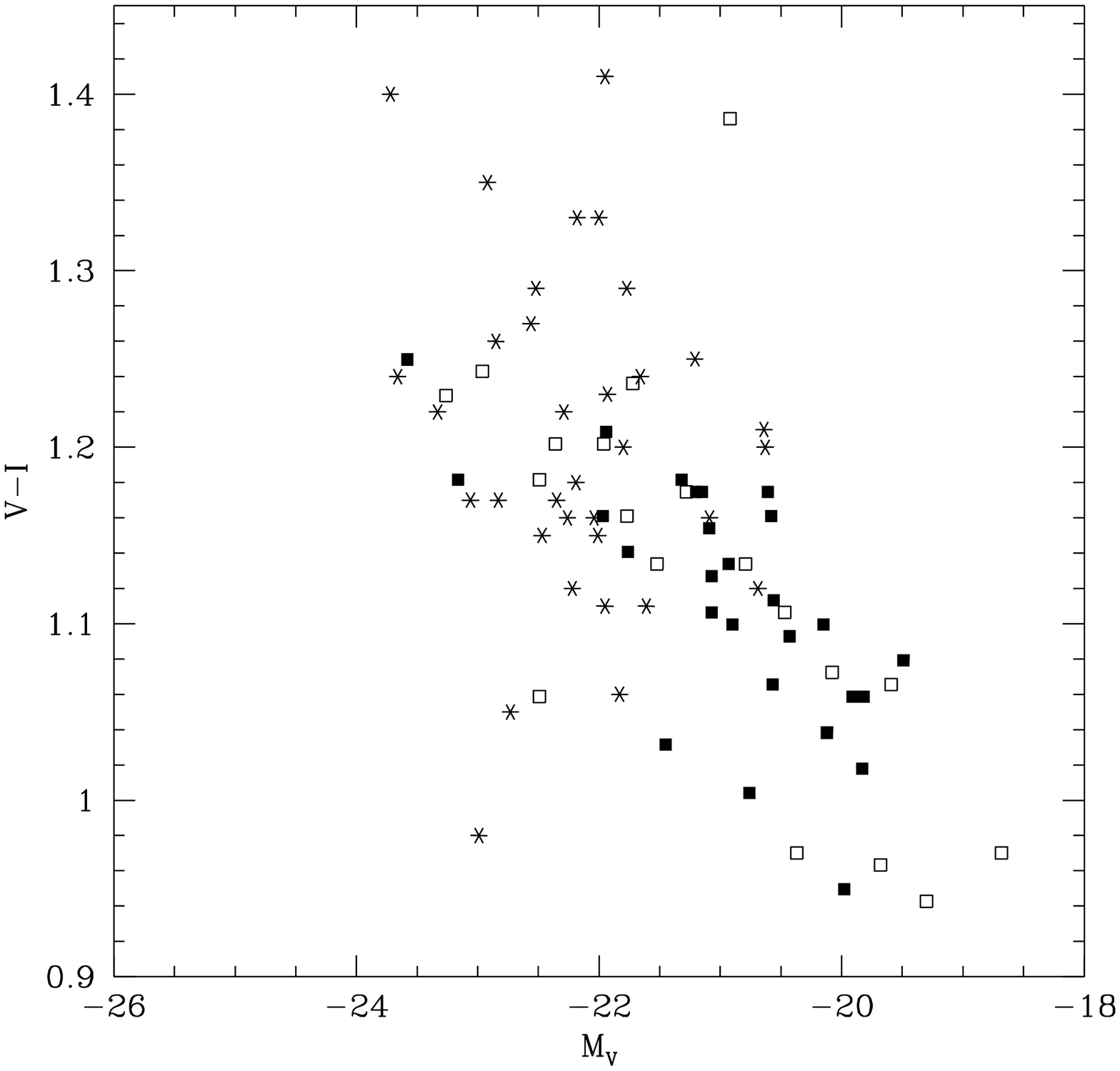}}
\caption{$V-I$ - $M_V$ colour-magnitude relation for the early-type
galaxies of Goudfrooij et al. (1994a) (asteriscs) together with the
data of Bower et al. (1992) converted to $V-I$ (data for Coma are shown
by filled squares and for Virgo by open squares). See text for more details}
\end{figure}

\twocolumn
~
\onecolumn

\appendix
\section{Colour transformations for the colour-colour diagram of 
the Coma cluster}

In this appendix we describe how a colour-colour relation for the Coma
cluster was obtained in $B-I$ and $I-H$. We started with the excellent
data of Bower et al. (1992) in $U$, $V$ and $K$. As a first attempt
we tried to make the colour conversions using single burst theoretical models
for old stellar populations using the Salpeter IMF. A fit to the models
of Vazdekis et al. (1996) gives

$$I-K ~~ = ~~ 0.546 ~ (V-K) ~ + ~ 0.275$$

and

$$B-V ~~ = ~~ 0.433 ~ (U-V) ~ + ~ 0.296$$

$I-K$ was finally converted to $I-H$ using the new GISSEL96 models 
of Bruzual \& Charlot (see Leitherer et al. 1996) using

$$I-H ~~ = ~~ 0.889 ~ (I-K) ~ + ~ 0.034$$

since the Vazdekis' models do not tabulate the $H$-band.
Alternatively, a fit to Worthey (1994)'s models gives:

$$I-K ~~ = ~~ 0.707 ~ (V-K) ~ - ~ 0.253$$

and

$$B-V ~~ = ~~ 0.407 ~ (U-V) ~ + ~ 0.379$$
 
using a very similar transformation from $I-K$ to $I-H$:

$$I-H ~~ = ~~ 0.895 ~ (I-K) ~ + ~ 0.047$$

There are several reasons to be wary of this approach. In the
first place one knows (e.g. Charlot et al. 1996) that the
infrared colours of the Worthey models for large metallicities
are too red by as much as a magnitude. Secondly, the model colours
of Vazdekis et al. in the infrared, just like those of Bruzual
\& Charlot (Leitherer {\it et al.} 1996) do not appear to be very accurate
either, since 
they show abrupt jumps as a function of age and metallicity,
due to the rather unknown contribution from stars in their late
stages of evolution. The conversion between $I-K$ and $I-H$
is only of minor importance. When one compares the results of both
methods, one finds that the difference
in the colour-colour diagram of Coma, when using two independent models,
is quite large (see Fig.~3c). This shows that a different, independent
way to convert the colours would be very useful, and for that reason
we attempted to derive an empirical conversion. 

In the search of data in $B$ or $I$ of the Coma cluster to combine
with Bower et al.'s dataset, we have found the data of J\o rgensen
{\it et al.}\  (1994), with photometry in $V$ and Gunn $r$, and the Ph.D. thesis
of 
Steel (1998), with $V$ and $R$. Since there is no $I$-band photometry
available, we decided to determine a conversion from $V-K$ to $V-R$,
a band with information very similar to $V-I$, and then convert $V-R$
to $V-I$ using published stars. It was also found that if one determines
a colour of a  galaxy using its integrated magnitudes, the observational
error would be much larger than if the same aperture was used both times.
For that reason we preferred to use the $V-R$ data inside 20$''$ of 
Steel (1998), for which the dispersion in the $V-R$ vs. $R_T$ was only
0.02 mag, rather than the total colours of J\o rgensen etal. (1994).
Combining the data of Steel (1998) with those of Bower et al. (1992)
one obtains 

$$V-R  ~~ = ~~ 0.336 ~ (V-K) ~ - ~ 0.501$$

with a scatter in $V-R$ of 0.03 mag. 
To convert $V-R$ to $V-I$ we fitted a least squares relation to all the 
Landolt (1992) standard stars. We find a very tight relation

$$V-I  ~~ = ~~ 2.029 ~ (V-R) ~ - ~ 0.018$$

with negligible scatter. The validity of this conversion between
$V-K$ and $V-I$ can be for example be established by looking at the 
$V-I$ - magnitude relation for the elliptical galaxies of Goudfrooij
{\it et al.} (1994a). This comparison gives satisfactory results.
Finally, we need to obtain a relation between
$U-V$ and $B-V$. Since here also there is quite a difference between 
the theoretical models, we have determined a least squares relation
between the observed $U-V$ and J\o rgensen's $B-r$, which we first 
converted to $B-V$ using the conversion given in J\o rgensen (1995):

$$B-V  ~~ = ~~ 0.673 ~ (B-r) ~ + ~ 0.184$$

to get:

$$B-V  ~~ = ~~ 0.356 ~ (U-V) ~ + ~ 0.448$$

Errors are difficult to determine. We can get an estimate by comparing
the data of Goudfrooij et al. (1994a) with the $V-K$ data of Bower et 
al. converted to $V-I$ (Figure 5). Plotted on the x-axis for the data
of Bower et al. are absolute $V$-band magnitudes, obtained using a distance 
modulus of 30.82 for Virgo and 34.51 for Coma (Aaronson et al. 1986), to which
a correction of 0.76 mag has been applied to convert them to integrated
magnitudes. For the data of Goudfrooij et al. (1994a) we plot their M$_V$ values.
There are three galaxies in common between the 2 samples, for which the 
M$_V$ values agree within 0.10 mag. Comparing the two samples we find that
for the galaxies between M$_V$ = --21 and --24 $V-I$ is redder by 0.032
mag on the average in the data of Goudfrooij et al. (1994a). This shows
that our conversion from $V-K$ to $V-I$ is probably reasonable, and based on this,
and on the tightness of individual
relations, we believe that the error in $V-I$ is at most 0.06 mag, the
same for $I-H$ or $I-K$, and 0.05 mag in $B-V$ , resulting in 0.08 mag in $B-I$.
We not only applied these transformation to the mean relation for the
Coma cluster, but also for the giant elliptical NGC 4472 (see Fig.~3).


\begin{thebibliography}{}
\bibitem[]{} Aaronson, M., Cohen, J.G., Mould, J. \& Malkan, M., 1978, 
ApJ, 223, 824
\bibitem[]{} Aaronson, M., Bothun, G., Mould, J., Huchra, J., 
Schommer, R.A. \& Cornell, M.E., 1986, ApJ, 302, 536
\bibitem[]{} Abraham, R.G., Ellis, R.S., Fabian, A.C., Tanvir, N.R.
\& Glazebrook, K., 1998, MNRAS, in press (astro/ph 9807140)
\bibitem[]{} Andredakis, Y., Peletier, R.F. \& Balcells, M., 1995, MNRAS, 275,
874
\bibitem[]{} Balcells, M. \& Peletier, R.F., 1994, AJ, 107, 135
\bibitem[]{} Baugh, C.M., Cole, S. \& Frenk, C.S., 1996, MNRAS, 282, L27
\bibitem[]{} Baugh, C.M., Cole, S., Frenk, C.S. \& Lacey, C.G., 1998,
ApJ, 498, 504
\bibitem[]{} Bender, R., Burstein, D. \& Faber, S.M., 1992, ApJ, 399, 462 
\bibitem[]{} Bouwens, R., Cayon, L. \& Silk, J., 1999, ApJ, in press 
\bibitem[]{} Bothun, G.D. \& Gregg, M.D., 1990, ApJ, 350, 73
\bibitem[]{} Bower, R.G., Lucey, J.R. \& Ellis, R.S., 1992, MNRAS, 254, 589
\bibitem[]{} Carlberg, R.G., 1984, ApJ, 286, 403
\bibitem[]{} Carollo, C.M., Franx, M., Illingworth, G.D. \& Forbes, D., 1997a,
ApJ, 481, 710
\bibitem[]{} Carollo, C.M., Stiavelli, M., de Zeeuw, P.T. \& Mack, J,
1997b, AJ, 114, 2366
\bibitem[]{} Carollo, C.M., Stiavelli, M. \& Mack, J., 1998, AJ, 116, 68
\bibitem[]{} Carter, D.,  1978, MNRAS, 182, 797
\bibitem[]{} Charlot, S., Worthey, G. \& Bressan, A., 1996, ApJ, 457, 625
\bibitem[]{} Colina, L., Holfeltz, S. \& Ritchie, C., 1998, in {\it 
NICMOS and the VLT}, ed. W.~Freudling and R.~Hook, 1998, ESO, p. 36
\bibitem[]{} Combes, F., Debbasch, F., Friedli, D., \& Pfenniger, D., 
1990, A\&A, 233, 82
\bibitem[]{} Davies, R.L., Efstathiou, G., Fall, S.M., Illingworth, G. 
\& Schechter, P.L., 1983, ApJ, 266, 41
\bibitem[]{}{} de Vaucouleurs, G., de Vaucouleurs, A., Corwin, H.G., Buta,
   R.J., Paturel, G., Fouqu\'e, P. 1991, 3rd Reference Catalogue of Bright 
   Galaxies (RC3), Springer, New York
\bibitem[]{} Djorgovski, S. \& Davis, M., 1987, ApJ, 313, 59
\bibitem[]{} Dressler, A., Lynden-Bell, D., Burstein, D., Davies, R.L.,
Faber, S.M., Terlevich, R. \& Wegner, G., 1987, ApJ, 313, 42
\bibitem[]{} Ebneter, K., Davis, M. \& Djorgovski, S., 1988, AJ, 95, 422
\bibitem[]{} Ellis, R., Smail, I., Dressler, A., et al., 1997, ApJ, 483, 582
{\it et al.}, ESA Publications (astro/ph 9808025)
\bibitem[]{} Faber, S.M. \& Gallagher, J.S. III, 1976, ApJ, 204, 365
\bibitem[]{} Garcia, A., A\& AS, 100, 47
\bibitem[]{} Goudfrooij, P., Hansen, L., J\o rgensen, H.E., N\o rgaard-Nielsen, 
H.U., de Jong, T \& van den Hoek, L.B., 1994a, A\&AS, 104, 179
\bibitem[]{} Goudfrooij, P., Hansen, L., J\o rgensen, H.E. \& N\o rgaard-Nielsen, 
H.U., 1994b, A\&AS, 105, 341
\bibitem[]{} Governato, F. et al., 1998, Nature, 392, 359
\bibitem[]{} Ho, L., Filippenko, A.V. \& Sargent, W.L.W., 1997, ApJS, 112, 315
\bibitem[]{} Hogg, D.W., astro-ph/9905116
\bibitem[]{} Holtzmann, J.A., Burrows, C.J., Casertano, S., Hester, J.J., 
Trauger, J.T., Watson, A.M. \& Worthey, G., 1995, PASP, 107, 1065
\bibitem[]{} Jaffe, W., Ford, H., Ferrarese, L., van den Bosch, F. \& O'Connell,
R.W., 1996, ApJ, 460, 214
\bibitem[]{} Kauffmann, G., 1995, MNRAS, 274, 153
\bibitem[]{} Kauffmann, G., 1996, MNRAS, 281, 487
\bibitem[]{} Knapen, J.H., Beckman, J.E., Shlosman, I., Peletier, R.F., Heller,
C.H. \& de Jong, R.S., 1995, ApJ, 443, L73
\bibitem[]{} Knapp, G.R., Guhathakurta, P., Kim, D.-W. \& Jura, M., 1989,
ApJs{70}, 329
\bibitem[]{} Kormendy, J. \& Illingworth, G.D., 1982, ApJ, 256, 460
\bibitem[]{} Krist, J. 1992, Tinytim v2.1 User's Manual (STScI) 
\bibitem[]{} Larson, R.B., 1975, MNRAS, 173, 671
\bibitem[]{} Leitherer, C., et al., 1996, PASP, 108, 996
\bibitem[]{} Lilly, S., et al., 1998, ApJ, 500, 75
\bibitem[]{} Madau, P., Ferguson, H.C., Dickinson, M. et al., 1996, MNRAS, 283,
1388
\bibitem[]{} Norman, C., Sellwood, J.A. \& Hasan, H., 1996, ApJ, 462, 114
\bibitem[]{} Ostriker, J.P., 1990, in {\it Evolution of the Universe 
of Galaxies}, ASP Conf. Ser. 10, ed. R. Kron (ASP, San Francisco), p. 25
\bibitem[]{} Peletier, R.F., Valentijn, E.A. \& Jameson, R.F., 1990, A\&A, 233, 62
\bibitem[]{} Peletier, R.F. \& Balcells, M., 1996, AJ, 111, 2238
\bibitem[]{} Peletier, R.F. \& Balcells, M., 1997, NewA, 1, 349
\bibitem[]{} Persson, S.E., Frogel, J.A. \& Aaronson, M., 1979, ApJs{39}, 61
\bibitem[]{} Pfenniger, D., 1992, in Proc. IAU Symp. 153, {\it Galactic
Bulges}, ed. H. Dejonghe \& H.J. Habing, Kluwer, Dordrecht, p. 387
\bibitem[]{} Pfenniger, D. \& Norman, C., 1990, ApJ, 363, 391
\bibitem[]{} Prada, F., Guti\'errez, C., Peletier, R.F. \& McKeith, C.D., 
1996, ApJ, 463, L9
\bibitem[]{} Rieke, G. \& Lebofsky, M.J., 1985, ApJ, 288, 618
\bibitem[]{} Sadler, E.M. \& Gerhard, O.E., 1985, MNRAS, 214, 177
\bibitem[]{} Sandage, A.R., Tammann, G., 1981, {\it A Revised Shapley-Ames
Catalog of Bright Galaxies}, Carnegie Institute of Washington
\bibitem[]{} Scalo, J.M., 1986, Fund. Cosmic Phys, 11, 1
\bibitem[]{} Schlegel, D.J, Finkbeiner, D.P. \& Davis, M., 1998, ApJ, in press
\bibitem[]{} Stanford, S.A., Eisenhardt, P.R. \& Dickinson, M., 1997,
ApJ, 492, 461
\bibitem[]{} Terndrup, D.M., Davies, R.L., Frogel, J.A., Depoy, D.L. 
\& Wells, L.A., 1994, ApJ, 432, 518
\bibitem[]{} van Dokkum, P.G. \& Franx, M., 1995, AJ, 110, 2027
\bibitem[]{} Vazdekis, A., Casuso, E., Peletier, R.F. \& Beckman, J.E., 1996,
ApJs, 106, 307
\bibitem[]{} V\'eron-Cetty, M.P. \& V\'eron, P., 1988, A\&A, 204, 28
\bibitem[]{} Wyse, R.F.G., Gilmore, G. \& Franx, M., 1997, ARAA, 35, 637
\bibitem[]{} Worthey, G., 1994, ApJs, 95, 107
\end{thebibliography}
\end{document}